\documentclass[conference]{IEEEtran}
\usepackage{color}
\usepackage{multirow}
\usepackage{cite}
\usepackage{balance}
\usepackage{amsmath}
\usepackage{amssymb}
\usepackage{enumitem}
\usepackage{arydshln}
\usepackage{fancyhdr}
\usepackage{marvosym}
\setlist[itemize]{noitemsep, topsep=0pt}
\usepackage{pdfsync}
\ifCLASSINFOpdf
\usepackage[pdftex]{graphicx}
\DeclareGraphicsExtensions{.pdf,.jpeg,.png}
\else
\usepackage[dvips]{graphicx}
\DeclareGraphicsExtensions{.eps}
\fi
\graphicspath{{figs/}}
\usepackage{epstopdf}
\usepackage{booktabs}
\usepackage{amsmath}

\renewcommand{\footnoterule}{%
  \kern -3pt
  \hrule width 0.49 \textwidth height 0.5pt
  \kern 1pt
}
\usepackage[absolute]{textpos}
\TPMargin{5pt}

\newcommand{\copyrightstatement}{
    \begin{textblock}{0.5}(0.08,0.94)   
    \noindent
         \footnotesize
          978-1-7281-0407-2/19/\$31.00~\copyright2019 IEEE 
    \end{textblock}
}

\begin{document}
\copyrightstatement
\title{Conservation Voltage Reduction by Coordinating Legacy Devices, Smart Inverters and Battery }

\author{\IEEEauthorblockN{Mohammad Ostadijafari, Rahul Ranjan Jha and Anamika Dubey}
\IEEEauthorblockA{School of Electrical Engineering and Computer Science \\
Washington State University\\
Pullman, WA}}


%


\maketitle

\begin{abstract}
Due to market, there is flexibility in the price for power purchase by the utility. At the maximum loading condition the cost of power is high, hence conservation voltage reduction (CVR) is a useful technique in reducing power demand of the distribution system. Further, by installing battery energy storage system (BESS) in the distribution system, power can be stored in the BESS when the price is low and can be utilize to supply the demand when the price is high. In this paper, power conserving and cost saving problems are investigated. As we consider loads in distribution system as voltage dependent loads, the CVR objective is obtained by optimizing the nodal voltages of a feeder towards the lower limit of the ANSI value using legacy devices and smart inverter. In order to solve introduced problems, a MPC-based algorithm is proposed for optimal usage of future information prediction such as load energy consumption profile, PVs generation profiles and TOU electricity tariffs. Moreover, effect of depreciation cost of the BESS is in cost saving is investigated. The proposed method is validated using IEEE-123 node system. It is demonstrated form the results that the proposed method is able to reduce the power consumption and cost of purchased power from substation.

\end{abstract}

\begin{IEEEkeywords}
		Conservation voltages reduction, Smart inverter, Distributed energy resources, Battery, Model predictive control.
\end{IEEEkeywords}
\vspace{0.4cm}

%
\IEEEpeerreviewmaketitle

\vspace{-0.45cm}

\section{Introduction}
Due to increase in the cost for power generation from the conventional source of energy, there is need for reduction in power consumption. Specially, at the peak load condition costly generators are used to provide the load demand. The generation cost are reduced by invoking the demand response technique in the distribution system. But, this technique only  tries to shift the peak load condition, it might be possible we will obtain another peak at different time of the day. Conservation voltage reduction (CVR) is a technique which is developed to reduce the energy consumption in the distribution system. The energy consumption for the voltage dependent loads can be reduced by decreasing the nodal voltages. It has been studied in \cite{schneider2010evaluation} that implementing CVR technique in all the feeders of the distribution system in United States it will cause 3.04\% reduction in annual energy consumption. 

The voltage control devices like capacitor banks (CBs), voltage regulators and on-load tap changers (OLTCs) can be used for controlling the voltage in the distribution system\cite{yeo2019power}. Earlier, CVR objective is obtained using these devices by measuring the feeder end voltage  such that all the nodes voltages are within the recommended ANSI voltage limits (0.95 - 1.05 pu)\cite{ANSI}. Now a days, a large amount of  distributed generators (DGs) are installed in the distribution system. These DGs are connected to the grid using smart inverters. According to  IEEE 1547-2018 standard \cite{IEEE1547},  these inverters are required to provide the reactive power support to  control the voltage in the system. The CVR objective is obtained by combining the legacy devices and smart inverters by solving the optimization problem called Volt-VAr Optimization (VVO).

The CVR objective can be obtained by formulating the problem as the optimal power flow (OPF) problem. A centralized controller is used to solve OPF by solving a mixed integer nonlinear problem (MINLP) for three-phase unbalanced system \cite{MINLPCVR,nojavan2014optimal,jabr2012minimum}. The problem is mixed integer due to inclusion of capacitor bank switch and tap position of the regulator. Unfortunately, solving MINLP for large scale three-phase unbalanced  distribution system is computationally expensive. Therefore, the nonlinear power flow equations in the distribution system are converted into linear power flow by either linearizing or approximating the equations\cite{paper1,LinearEmiliano1, gan2014convex,LinearEmiliano2}. 


Recently, photovoltaic (PV) generators \cite{teymouri2015advanced} are installed in the distribution system, which are interfaced with battery energy storage system (BESS)\cite{ostadijafari2019smart}. The main purpose of BESS is to purchase and store  low price energy during off-peak times (charging state) and utilize it during peak times when the energy price is high (discharging state)\cite{mohammadi2016allocation}. For example, \cite{rahimi2013simple} and \cite{barnes2012placement} use BESS to perform peak load shaving which provides direct and indirect benefits such as cost and/or energy savings and voltage support. In \cite{yeh2012performance} and \cite{yeh2013battery}, two dynamic control methods for reactive power control in a single line radial distribution system based on BESS and PV's  inverters are proposed and compared. In addition, advanced control methods such as model predictive control (MPC) can be exploited in this context to obtain most benefits from BESS. In \cite{zafar2018multi}, author proposes a MPC-based optimization to coordinate the substation OLTC operation on slow-timescale with  inverters and BESS on fast-timescale. Then, this multi-timescale volt/var optimization problem is formulated as the mixed-integer second-order cone program (MISOCP) with objective of minimizing the power loss and total energy cost of the system. In this work we propose a MPC-based optimization to coordinate the legacy devices, smart inverters and BESS. The problem is formulated as MILP with CVR objective and minimizing total cost of purchasing energy from grid given the day-ahead time-of-use (TOU) electricity tariff, depreciation cost of the BESS, forecasted load demand and PV profiles. 

The rest of this paper is as follows: Section \ref{model} provides the modeling of the distribution system with voltage control devices and BESS. In section \ref{Sec III}, problems of interest, and a MPC-based algorithm for solving them are explained. Section \ref{simulation} provide the discussion and the results on IEEE-123 node system. Finally, Section \ref{conclusion} provides concluding of the paper.
 \vspace{-0.2cm}
\section{Modelling of Distribution System }
\label{model}\vspace{-0.2cm}
In this section, first, we introduce the branch-flow model for the distribution system. Then a brief introduction on the modelling of all the  components in the distribution system is provided for the optimization purpose. 
 \vspace{-0.35cm}\subsection{Branch Flow Model} 
\vspace{-0.2cm}
\begin{figure}[t]
\vspace{-0.3cm}
\centering
\includegraphics[width=2.7in]{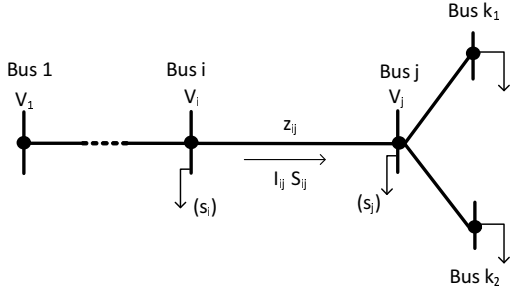}
\vspace{-0.5cm}
	\caption{Topology of radial distribution system }
	\vspace{-0.7cm}
	\label{fig:1}
	\end{figure}

The distribution system can be modelled as a directed graph as shown in Fig.1. Consider a distribution system consisting  of $\mathcal{N}$ number of buses and $\mathcal{E}$  number of edges. As most of the distribution system operates radially, hence, each  edge ($i,j$) connects nodes $i$ and $j$ where node $i$ is the parent of  node $j$. The impedance of an edge is represented as $z_{ij} = r_{ij} + \iota x_{ij}$, the power flowing in the edge is given as $S_{ij} = P_{ij} + \iota Q_{ij}$ and the complex line current is $I_{ij}$. The voltage of a node $i \in \mathcal{N}$ is given by $V_i$ (a complex number) and $s_i = p_i + \iota q_i$ be the net apparent power injection (generation minus demand) at corresponding bus $i$. The the branch-flow equations for radial distribution feeder,  is given in (1)-(4). Please refer to \cite{MFarivarLow} for further details.

\begin{small}\vspace{-0.5cm}
\begin{eqnarray}
p_j &=& P_{ij}  -  \sum_{k:j\to k}P_{jk} + r_{ij}l_{ij}  \hspace{2 cm} \forall i \in \mathcal{N}\\
q_j  &=& Q_{ij} - \sum_{k:j\to k}Q_{jk} + x_{ij}l_{ij}  \hspace{2 cm} \forall i \in \mathcal{N}\\
v_i &=& v_j + 2(r_{ij}P_{ij} + x_{ij}Q_{ij}) - (r_{ij}^2 + x_{ij}^2)l_{ij}  \forall i \in \mathcal{N}\\
v_il_{ij} &=& P_{ij}^2 + Q_{ij}^2   \hspace{4 cm} \forall {ij} \in \mathcal{E}
\end{eqnarray}
\vspace{-0.5cm}\end{small}

Note, that the branch flow equations in (1)-(4) are obtained by relaxing the nodal voltage angles as described in \cite{MFarivarLow}. For a radial system, the nodal voltage angle can be exactly obtained from the OPF result. Also, notice that $v_i = |V_i|^2$ and $l_{ij} = |I_{ij}|^2$. The power flow equation described in equation (1)-(4) is nonlinear due to $l_{ij}$. It is known that the power flowing in  each edges of the distribution system is much larger than the losses incurred in these edges. Hence, in order to obtain the linear power flow equation, $l_{ij}$ term is ignored  to obtain the linear set of equations as given below:

\begin{small}\vspace{-0.3cm}
\begin{eqnarray}
p_j &=& P_{ij}  -  \sum_{k:j\to k}P_{jk}   \hspace{3.1 cm} \forall i \in \mathcal{N}\\
 q_j  &=& Q_{ij} - \sum_{k:j\to k}Q_{jk}   \hspace{3.1 cm} \forall i \in \mathcal{N}\\
v_i &=& v_j + 2(r_{ij}P_{ij} + x_{ij}Q_{ij}) \hspace{2 cm} \forall i \in \mathcal{N}
\end{eqnarray}
\vspace{-0.05cm}\end{small}
The linear power flow equations in (5)-(7) closely approximate the power flow model for the unbalanced three-phase system which is validated in our previous work. Please refer to \cite{paper1} where the results shows that the voltage obtained using linear sets of equations have error in the range of $0.001$ pu compared to actual power flow results. 
 \vspace{-0.1cm}
\subsection{Distribution System Components}
\subsubsection{Voltage Regulators}  
The voltage regulation range of a voltage regulator is assumed to be $\pm10\%$, which is divided into 32-steps for this work. The series and shunt impedance of the  voltage regulator have very small values and are ignored for simplification. The primary and secondary nodal voltages for the voltage regulator is related to  $a$ of the regulator by:
 \vspace{-0.2cm}\begin{eqnarray} \label{eq11} \vspace{-0.3cm}
V_j = a  V_i 
\end{eqnarray}
where, $a = \sum\limits_{i=1}^{32} b_i u_{tap,i}$ and  $\sum\limits_{i=1}^{32} u_{tap,i} = 1$. As square of voltage magnitude is the variables in our system hence, voltage relation for regulators are in terms of $v_i$ and $v_j$ and  $A$ = $a^2$.
 \vspace{-0.3cm}
\begin{eqnarray} \label{eq12}
v_j = A \times v_i
\end{eqnarray}


\vspace{-0.1cm}
\subsubsection{Capacitor Banks}
The reactive power $q_{cap,i}$ provided by the capacitor bank are function of square of voltages and the switch status (On/Off).
\vspace{-0.1cm}\begin{equation} \label{eq13}\small
q_{cap,i} = u_{cap,i} q_{cap,i}^{rated} v_{i}
\end{equation}where, $u_{cap,i}$ is the switch status, $q_{cap,i}^{rated}$ rated power of the capacitor bank and $v_{i}$ is the square of the bus voltage.


 \subsubsection{Distributed Generation with Smart Inverters}
The DGs are modeled as negative loads with a known active power generation $p_{DG,i}$ equal to the forecasted value. The reactive power support from the DGs depends on the rating of the smart inverter. In this work, the smart inverter is rated at $ 15 \% $ higher than the maximum active power rating of the DGs. 
 \begin{equation} \label{eq14}
\small
 -\sqrt{(s_{DG,i}^{rated})^2 - (p_{DG,i})^2} \leq q_{DG,i} \leq \sqrt{(s_{DG,i}^{rated})^2 - (p_{DG,i})^2}
\end{equation}where, $s_{DG,i}^{rated}$ is the apparent power of DG, $q_{DG,i}$ reactive power support from smart inverters. 

\subsubsection{Voltage-Dependent Load Models}

The most widely acceptable load model is the ZIP model which is a combination of constant impedance (Z), constant current (I) and constant power (P)) characteristics of the load \cite{kers}. In this paper a voltage dependent load model, which is function of CVR coefficients is used as given by (12)-(13).
\begin{small}\begin{eqnarray} \label{eq18}
p_{L,i}^p = p_{i,0}^p + CVR_{p}\dfrac{p_{i,0}^p}{2}\left(\dfrac{v_i^p}{V_0^2}-1\right)\\
q_{L,i}^p = q_{i,0}^p + CVR_{q}\dfrac{q_{i,0}^p}{2}\left(\dfrac{v_i^p}{V_0^2}-1\right)
\end{eqnarray}\end{small}
Here, the CVR coefficients are obtained from the ZIP coefficients. It has been observed that the proposed load model has similar characteristics of ZIP model in the voltage range of 0.95-1.05 pu. The detailed derivation for the proposed load model can be found in \cite{paper1}.

\subsection{BESS Model}
\label{Battery storage model}
The BESS dynamics can be presented based on state-of-charge (SOC) and the limits on the BESS charging and discharging power and energy\cite{wei2016proactive}. This is formulated as the  following: 
\vspace{-0.1cm}
    \begin{equation}\label{eq-bat1}
    \small
	SOC(t+1)=(1-\eta)SOC(t)+\rho \frac{P_{c,d}(t)}{Q_{bat}}\tau
	\end{equation}
	\vspace{-0.3cm}
	\begin{equation}\label{eq-bat2}
	\small
	E^{-}\leq SOC(t+1) \leq E^{+}
	\end{equation}
	\vspace{-0.3cm}
	\begin{equation}\label{eq-bat3}
	\small
	-d_r\leq P_{c,d}(t) \leq c_r 
	\end{equation}
	\vspace{-0.3cm}\begin{equation}\label{eq-newbattery}
	\small
	P_d(t)=\begin{cases}
	|P_{c,d}(t)| & \text{if $P_{c,d}(t)<0$}\\
	0 & \text{otherwise}
	\end{cases}\end{equation}
The BESS SOC updates is given in (\ref{eq-bat1}) where $SOC(t)$ and $P_{c,d}(t)$ are the SOC and charging/discharging power of the BESS at sampling time $t$, respectively. $\eta$, $\rho$, and $Q_{bat}$ are energy decay rate, round-trip efficiency, capacity of the BESS and $\tau$ is the sampling time-interval. Constraint (\ref{eq-bat2}) guarantees the BESS SOC remains within a boundary where $E^{+}$ and $E^{-}$ specifies bounds on BESS charging/discharging limits. Finally, constraint (\ref{eq-bat3}) bounds BESS maximum charging/discharging rates where $d_r$ is the maximum discharge rate while $c_r$ is the maximum charge rate. Absolute value of the BESS discharging power is calculated in (\ref{eq-newbattery}) and denoted by $P_d(t)$. Also,  $P_{c,d}(t)>0$ indicates that BESS is charging while $P_{c,d}(t)<0$ indicates that the BESS is discharging. The current formulation for the BESS is nonlinear and adds computational complexity to the problem. 
Note that constraint (\ref{eq-newbattery}) makes BESS model formulation nonlinear. Same as the nonlinear power flow equations, this also can add computational complexity to the optimization problem. Therefore we linearize the constraint (\ref{eq-newbattery}) as follows:
    \begin{equation}\label{eq-app1}
	\small
	P_d(t)=\begin{cases}
	|P_{c,d}(t)| & \text{if $-P_{c,d}(t) > 0$}\\
	0 & \text{otherwise}
	\end{cases}\end{equation}
Different conditions of (\ref{eq-app1}) can be modeled by using a binary variable and Big-M method. First, a binary variable is defined to activate different boundaries of (\ref{eq-app1}). This can be stated as follows: 
\vspace{-0.3cm}
\begin{small}
\begin{eqnarray}\label{eq-app2}
           \delta=1 \Longleftrightarrow -P_{c,d}(t) > 0  \\
\nonumber  \delta=0 \Longleftrightarrow -P_{c,d}(t) \leq 0 
	\end{eqnarray}
	\end{small}This can be formulated as:
	\begin{small}
	\begin{equation}
	\label{eq-app3}
-P_{c,d}(t)\geq \epsilon -M(1-\delta) 
	\end{equation}
	\end{small}
	\vspace{-0.3cm}
		\begin{small}
	\begin{equation}
	\label{eq-app4}
	-P_{c,d}(t)\leq M\delta
	\end{equation}\end{small}Where, $\delta \in \{0,1\}$ is binary variable; $M$ is an arbitrary large number and should be chosen as a big enough (possibly always different) number to not limit any significant function of variables in the constraints; Also, $\epsilon$ is an arbitrary small number to guarantee that for $\delta=1$, $P_{c,d}(t)$ is not equal to zero. Then, the conditions of (\ref{eq-app1}) can be modelled based on (\ref{eq-app2}) as follows:
\vspace{-0.32cm}
\begin{small}
\begin{eqnarray}\label{eq-app5}
\delta=1 \Longleftrightarrow P_d(t)=|P_{c,d}(t)| \\
\nonumber\delta=0 \Longleftrightarrow P_d(t)=0 \ \quad
	\end{eqnarray}
	\end{small}
 \vspace{-0.15cm}This can be formulated as:
\begin{equation}
	\small
	    \label{eq-app6}
	    |P_{c,d}(t)|-M(1-\delta)\leq P_d(t) \leq |P_{c,d}(t)|+M(1-\delta)
	\end{equation}
		\begin{equation}
	\small
	    \label{eq-app7}
	    -M\delta\leq P_d(t) \leq M\delta
	\end{equation}Due to the absolute value term in the double inequality (\ref{eq-app6}), this constraint still needs to be linearized. Thus, the left hand-side inequality of (\ref{eq-app6}) can be modified to write as follows:

\vspace{-0.3cm}
\begin{equation}
	\small
	    \label{eq-app8}
	    |P_{c,d}(t)|\leq P_d(t)+M(1-\delta)
	    \vspace{-0.1cm}
	\end{equation}
which can be linearized as follows:
\vspace{-0.2cm}
	\begin{equation}
	\small
	    \label{eq-app9}
	    P_{c,d}(t)\leq P_d(t)+M(1-\delta)
	\end{equation}
		\begin{equation}
	\small
	    \label{eq-app10}
	    -P_{c,d}(t)\leq P_d(t)+M(1-\delta)
	\end{equation}
	Similarly, the right hand-side inequality of (\ref{eq-app6}) can be stated as follows:

\vspace{-0.3cm}
\begin{equation}
	\small
	    \label{eq-app11}
	    |P_{c,d}(t)|\geq P_d(t)-M(1-\delta)
	\end{equation}
which can be linearized as follows:
					\begin{equation}
	\small
	    \label{eq-app12}
	    P_{c,d}(t)\geq P_d(t)-M(1-\delta)-M\beta
	\end{equation}
						\begin{equation}
	\small
	    \label{eq-app13}\vspace{-0.1cm}
	    -P_{c,d}(t)\geq P_d(t)-M(1-\delta)-M(1-\beta)
	\end{equation}
	where, $\beta \in \{0,1\}$ is a binary variable which is defined to linearize (\ref{eq-app11}). Therefore, the nonlinear constraint (\ref{eq-newbattery}) can be replaced by set of mixed-integer constraints (\ref{eq-app3}), (\ref{eq-app4}), (\ref{eq-app7}), (\ref{eq-app9}), (\ref{eq-app10}), (\ref{eq-app12}) and (\ref{eq-app13}).
	\vspace{-0.1cm}
\section{Predictive Controller}
\label{Sec III}\vspace{-0.1cm}
In this work, it is assumed that utility is equipped with an intelligent controller which has the access to the day-ahead forecast of the distribution system's load energy consumption profile, PV generation profile and TOU electricity tariffs. In order to satisfy the distribution system energy demand, the utility intelligent controller can provide electricity from any combination of PV generation, BESS, and electricity purchased from the retail electricity provider (here we refer  it as substation). 
The intelligent controller at utility optimally co-schedules BESS and available energy resources based on the desirable objective of the utility. In this paper, we consider two different objective functions for the utility intelligent controller based on energy management and revenue management purposes. In both cases, the problem formulated as the multi-time optimization problem which can be regarded as the MPC-based formulation. In what follows, each of these cases is explained thoroughly.

\subsection{Energy management}
\label{Energy management}
The energy management problem is formulated as an MPC-based problem with the objective of meeting load demands while minimizing the total purchased energy from substation by co-scheduling legacy devices, smart inverters, and BESS. In this problem, the voltage in the distribution system is controlled towards the lower range of ANSI value. As the loads in the system are modelled as voltage dependent loads, this will lead to minimization of active power consumption of the system at every 15-min time interval for a day. 
The energy minimization problem is formulated as shown below:

\begin{small}
\begin{equation}\label{eq36}
\underset{U}{Min} \sum_{t=m}^{m+W-1} P_T(t) 
\end{equation}
\begin{equation*}
U= \{u_{tap,i}(t),u_{cap,i}(t),q_{DG,i}(t),P_{c,d}(t)\}\hspace{0.1cm} \text{and}\hspace{0.1cm}\forall t\in [m,t+m+1]
\end{equation*}
\end{small}
Subject to:
\begin{small}
\begin{flalign}\label{eq23}
 &P_{ij}(t) = \sum_{k:j \rightarrow k}P_{jk}(t) + p_{L,j}(t)- p_{DG,i}(t) \hspace{1.15cm} \forall i\in\mathcal{N} \\
&Q_{ij}(t) = \sum_{k:j \rightarrow k}Q_{jk}(t) + q_{L,j}(t)- q_{DG,i}(t) - q_{C,i}(t) \hspace{0.2cm} \forall i\in\mathcal{N} \\
            &v_i(t) = v_j(t) + 2(r_{ij}P_{ij}(t) + x_{ij}Q_{ij}(t))   \hspace{1.55cm} \forall j \in \mathcal{N}  \\
            &p_{L,i}(t) = p_{i,0}(t) + CVR_{p}(t)\dfrac{p_{i,0}(t)}{2}(v_i(t)-1) \hspace{0.57cm} \forall i\in\mathcal{N_L} \\
            &q_{L,i}(t) = q_{i,0}(t) + CVR_{q}(t)\dfrac{q_{i,0}(t)}{2}(v_i(t)-1) \hspace{0.62cm} \forall i\in\mathcal{N_L} \\
            &v_j(t) = A_i(t) v_i(t) \hspace{3.9cm} \forall (i,j) \in \mathcal{E_T} \\
            &A_i(t) = \sum\limits_{i=1}^{32} B_i u_{tap,i}(t), \sum\limits_{i=1}^{32} u_{tap,i}(t) = 1 \hspace{0.8cm} \forall (i,j) \in \mathcal{E_T}  \\
            &q_{C,i}(t) = u_{cap,i}(t)  q_{cap,i}^{rated} v_i(t)  \hspace{2.7cm} \forall (i) \in \mathcal{N_C} \\
            &q_{DG,i}(t) \leq \sqrt{(s_{DG,i}^{rated})^2 - (p_{DG,i}(t))^2} \hspace{1.37cm} \forall (i) \in \mathcal{N_{DG}} \\
            & q_{DG,i}(t) \geq -\sqrt{(s_{DG,i}^{rated})^2 - (p_{DG,i}(t))^2} \hspace{1.1cm} \forall (i) \in \mathcal{N_{DG}}\\
            &(0.95)^2\leq v_i(t) \leq (1.05)^2 \hspace{3.5 cm} \forall i\in \mathcal{N}\\
            &P_T(t)=P_s(t)+P_{c,d}(t)
 \end{flalign}\end{small}
	\begin{small}
	\vspace{-0.5cm}
	\begin{align*}
	\text{and BESS constraints (\ref{eq-bat1})- (\ref{eq-bat3})  }
	\end{align*}
	\end{small}where (\ref{eq36}) indicates that the objective of the problem is to minimize the total purchased power from substation ($P_T(t)$) during the prediction window $W$, by scheduling BESS, capacitor bank, regulator tap position and reactive power of  inverter at different sampling times of a day. Equations (32), (33) and (34) are the power flow and voltage constraints in the distribution system. The voltage dependent loads are added as constraints  in equation (35) and (36) for the active and reactive power loads, respectively. The capacitor banks and voltage regulators constraints are given by equation (37), (38) and (39). The lower and upper limits on the reactive power support by smart inverters are given by (40) and (41). The lower and upper limits on nodal voltages are  given by equation (42).  Constraint (43) states that the total purchased power from substation at each sampling time is equal to summation of required power of the loads and the charging/discharging power of the BESS. Note that (\ref{eq-newbattery})-(\ref{eq-app13}) are not mentioned in BESS constraints of the objective (\ref{eq36}). The reason is that as the objective of the problem is defined based on minimizing the purchased power from the substation, the intelligent controller of the utility schedules the BESS regardless of it's depreciation cost; thus, constraints (\ref{eq-newbattery})-(\ref{eq-app13}) which are used in BESS model for considering depreciation cost do not change the answer of the problem. 
\subsection{Revenue management}
\label{Revenue management}
The revenue management is formulated as a MPC-based problem with the objective of meeting load demands while minimizing the total cost of purchasing energy. The emphasis in this problem is to optimally co-schedule legacy devices, smart inverter and BESS to minimize total purchase energy cost during the day. The day ahead-information include a price vector whose entities are TOU electricity tariffs of the day-ahead in addition to the forecast of energy consumption and generation. This can be formulated as follows:

\begin{small}
\begin{equation}\label{eq49}
\underset{U}{Min} \sum_{t=m}^{m+W-1} Price(t).P_T(t)+{Price}_{b} .P_{d}(t)
\end{equation}
\begin{equation*}
U= \{u_{tap,i}(t),u_{cap,i}(t),q_{DG,i}(t),P_{c,d}(t)\}\hspace{0.1cm} \text{and}\hspace{0.1cm}\forall t\in [m,t+m+1]
\end{equation*}
\end{small}
Subject to:
	\begin{small}
	\vspace{-0.1cm}
	\begin{align*}
	\text{Constraints (\ref{eq23})- (42) and BESS constraints (\ref{eq-bat1})- (\ref{eq-app13})  }
	\end{align*}
		\end{small}
\vspace{-0.5cm}

The minimization of the purchased electricity cost from substation is given by the first term of (\ref{eq49}) while the second term minimizes the BESS depreciation cost. In (\ref{eq49}), ${Price}_{b}$ is the BESS depreciation cost and $Price(t)$ the electricity tariff at the sampling time $t$, respectively. As it can be seen all the constraints of the revenue management problem are same as the energy management problem except (\ref{eq-newbattery})-(\ref{eq-app13}) which are used in BESS model for considering depreciation cost due to the cost minimization objective of revenue management problem.

The depreciation cost of the BESS  is an index to show how fast the BESS life time is degrading. This helps the utility controller to determine if it is economical to utilize the BESS in each sampling time.
\subsection{MPC Modelling}
\label{MPC}
 At the beginning of each day, utility intelligent controller receives one-day ahead prediction information including load energy consumption and PV generation profiles. Note that for the case of revenue management problem formulated in Section \ref{Revenue management}, the controller also needs to access TOU electricity tariffs for next 24 hours of the day. The MPC-based algorithm solves the minimization problems (\ref{eq36}) and (\ref{eq49}), with their corresponding constraints described in sections \ref{Energy management} and \ref{Revenue management}, respectively. This results in set of trajectories\begin{small} $[u_{tap,i}(t),u_{tap,i}(t+1), ...,u_{tap,i}(t+W-1)]$, $[u_{cap,i}(t),u_{cap,i}(t+1), ...,u_{cap,i}(t+W-1)]$, $[q_{DG,i}(t),q_{DG,i}(t+1), ...,q_{DG,i}(t+W-1)]$\end{small} and \begin{small}$[{P}_{c,d}(t),{P}_{c,d}(t+1), ...,P_{c,d}(t+W-1)]$\end{small} which indicate capacitor bank, regulator tap position and reactive power of smart inverter of bus $i$ and charging/discharging of BESS trajectories for a prediction window from time \begin{small}$t$\end{small} to time \begin{small}$t+W-1$\end{small}. After obtaining the optimal set of trajectories, only the first entry of these trajectories \begin{small}($u_{tap,i}(t)$, $u_{cap,i}(t)$, $q_{DG,i}(t)$\end{small} and \begin{small}${P}_{c,d}(t)$)\end{small} are applied for controlling the legacy devices, smart inverter and  BESS operation. Then, the MPC-based algorithm moves one step forward; prediction window changes accordingly (from \begin{small}$t+1$\end{small} to \begin{small}$t+W$\end{small}) and MPC algorithm repeated again. 
 
\begin{figure}[t]
		\centering
		\vspace{-0.2cm}
		\includegraphics[width=0.95\linewidth]{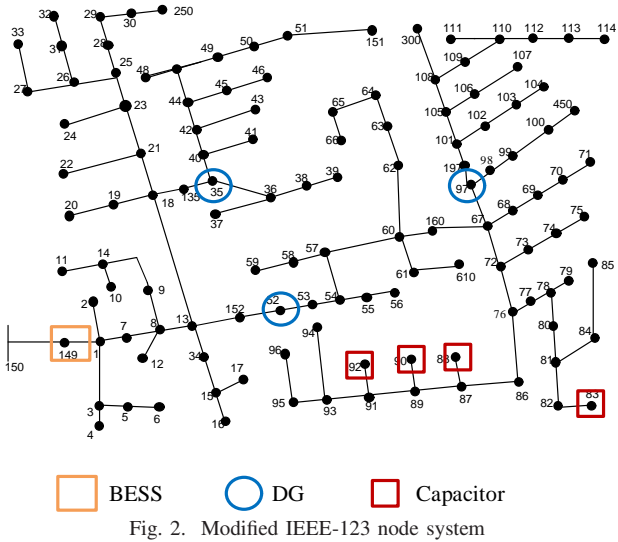}
	\vspace{-0.4cm}
		\caption{Modified IEEE-123 node system }
		\label{fig:123}
		\vspace{-0.4cm}
\end{figure}

\begin{figure}[t]
		\centering
		\includegraphics[width=\linewidth]{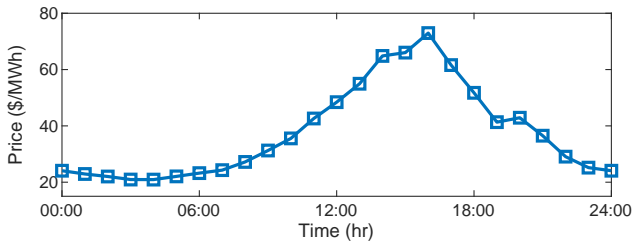}
	\vspace{-0.7cm}
		\caption{TOU electricity tariffs}
		\label{fig:price}
		\vspace{-0.6cm}
	\end{figure}
	
\section{Simulation Results}
 \vspace{-0.1cm}
\label{simulation}
In this section, we conduct a set of simulation studies to validate the performance of the proposed MPC-based algorithm for solving the revenue and energy management problems introduced in Section \ref{Sec III}. Next, we focus on the role of choosing the BESS in cost saving problem based on depreciation cost of the BESS. 
Simulations are performed on IEEE-123 bus system (see Fig. \ref{fig:123}). 
 Equivalent single phase model of IEEE-123 node system is modified by adding three PVs of 115 kVA rating each at node 35, 52 and 97. In both systems, the BESS with following parameters is used: \begin{small}$\eta=0,\rho=1, E^{-}=0.25, E^{+}=1, d_r=c_r=100$ kW\end{small} and \begin{small} $Q_{bat}=300 $ kWh\end{small}. The CVR coefficients for active and reactive power load are taken as 0.6 and 3 respectively. The load demand curve and the PV generation profile for a day for every 15-min time interval are obtained from the OpenDSS. In addition, Fig. \ref{fig:price} shows the 24-hour TOU electricity tariffs used in simulations.


\subsection{Effect of MPC control on energy and revenue management}
\label{Sec:IVA}
\begin{figure}[t]
		\centering
		\vspace{-0.2cm}
		\includegraphics[width=\linewidth]{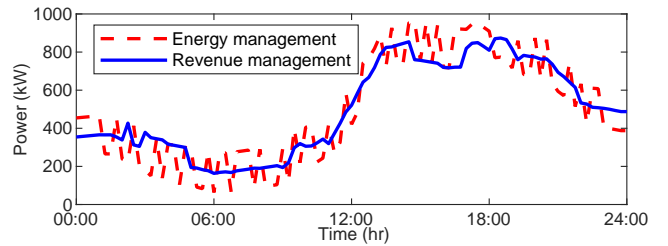}
	\vspace{-0.8cm}
		\caption{Total purchased electrical power from substation during a day based on defined problems}
		\label{fig:P}
		\vspace{-1.8cm}
	\end{figure}

The comparison of the energy and revenue management problems based on electrical power purchased by utility from substation is shown for IEEE-123 node system in Fig. \ref{fig:P}. It can be observed that during times with low TOU electricity tariffs (e.g. 04:00-06:00), the optimal solution for the utility controller with revenue management objective is to purchase more power from the substation compare to the energy management objective. That is, the focus of revenue management problem for the utility is cost saving; thus, utility purchases the cheap energy at these times and utilize BESS to store it. The stored energy is injected back to the distribution system at the times with high TOU electricity tariffs (e.g. 15:00-17:00). Hence, less power with expensive price is purchased from the substation compared to the other objective.  However, in the event of utility controller with energy management objective, the total power purchased from substation during the day is less than the case with revenue management objective. Note that in the case of utility controller with energy management objective, BESS  charges and discharges regardless of the TOU electricity tariffs to reduce nodal voltages for CVR purpose. 

This can be verified in Table \ref{table:emad1} where the cost and amount of purchased power from the substation during the day are compared for the both objectives. As it can be seen, using revenue management objective for utility controller leads to decreasing the total cost of purchased electrical power from substation during a day while the utility controller with energy management objective purchases less amount of electrical power during a day.
\begin{table}[ht]
\vspace{-0.4cm}
\centering
\caption{cost and amount of purchased power from the substation during the day for defined objective}
\vspace{-0.2cm}
\label{table:emad1}
\begin{tabular}{c|c|c}
                  & Cost (\$) & Power (kW) \\ \hline\hline
Energy Management &  2186         & 48130           \\ \hline
Revenue Management   &      2132     &  48132          \\ \hline\hline
\end{tabular}
\vspace{-0.13cm}
\end{table}
	
Next, the effect of VVO with CVR objective is simulated on IEEE-123 node system. The effect of legacy devices are shown for the maximum and minimum loading condition for both the energy management and revenue management case. At the minimum loading condition, the voltage regulator tap is at -16 and all the capacitor switch are OFF, the required power supplied by the DGs is shown in Table II. Further, at maximum loading condition, The tap position is at -12 and  three-capacitor banks is ON and one is OFF to provide the reactive power support. The DGs are supplying/absorbing reactive power accordingly to maintain the nodal voltages within the limits. It can be observed from table II, that for both the energy management and the revenue management the VVO is achieving the same CVR objective, so the control status in both the cases are same.
\vspace{-0.1cm}
\begin{table}[ht]
\centering
\caption{Volt-VAr control for IEEE-123}
\vspace{-0.2cm}
\begin{tabular}{c|c|c|c}
\hline
\multicolumn{4}{c}{Energy Management }\\
\hline
loading condition & $u_{tap}$ & $u_{cap}$ & $q_{DG}$(kVAr) \\
\hline
minimum & -16 & 0,0,0,0 & 42.6,67.2,73.9 \\
\hline
maximum & -12 & 1,1,1,0& -15.6,55.2,49.6\\
\hline
\multicolumn{4}{c}{Revenue Management }\\
\hline
loading condition & $u_{tap}$ & $u_{cap}$ & $q_{DG}$(kVAr) \\
\hline
minimum & -16 & 0,0,0,0 & 42.6,67.2,73.9 \\
\hline
maximum & -12 & 1,1,1,0& -15.6,55.2,49.6\\
\hline
\hline
\end{tabular}
\vspace{-0.2cm}
\end{table}

\subsection{Depreciation cost of the BESS}
\label{Sec:IVB}
\begin{figure}[t]
		\centering
		\vspace{-0.3cm}
		\includegraphics[width=\linewidth]{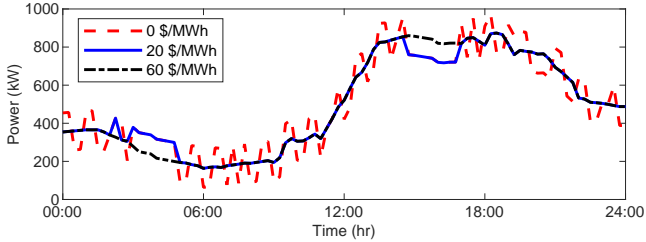}
	\vspace{-0.8cm}
		\caption{Amount of purchased power during a day based on different depreciation costs of the BESS }
		\label{fig:deP}
		\vspace{-0.8cm}
	\end{figure}

  In this section, we investigate the effects of depreciation cost of the BESS in the case of energy management objective on cost saving. Fig. \ref{fig:deP} shows the purchased power from substation during the day for three different deprecation costs of the BESS. As it can be seen, when ${Price}_{b}=0$, the intelligent utility controller charges and discharges the BESS without any limitation and only based on TOU electricity tariffs. Thus, there are more oscillations due to charging/discharging in the utility purchased power trajectories compared to the other cases with different depreciation costs. This is also the case for all the depreciation costs less than the minimum predicted TOU electricity tariffs of a day which allows the most possible utilization of BESS.  As the depreciation cost of the BESS increases, it will be less economical for utility to use the BESS. For example when ${Price}_{b}=20$, the BESS charges (discharges) at minimum (maximum) TOU electricity tariffs. 
  At the extreme case when ${Price}_{b}=60$, the BESS does not charge/discharge in any sampling time. This is also the case for all the depreciation costs higher than the maximum predicted TOU electricity tariffs of a day which is equivalent with the case the utility does not posses BESS. 
\vspace{-0.2cm}	
\section{Conclusion}
\label{conclusion}
Conservation of electrical power and cost saving in purchasing energy are important concerns for the today's power market. In this work, an MPC-based algorithm for energy and revenue management of the distribution utility was presented.  The aim in energy management problem is to conserve electrical power while in revenue management problem, the aim is to reduce the cost of purchased electrical power from retail electricity provider. In both problems, MPC-based algorithm controls the legacy devices, smart inverters and BESS to meet desired objective of utilities. 
Further, the effect of choosing BESS based on it's depreciation cost was investigated. The  proposed algorithm was validated by using IEEE-123 node system and efficiency of the algorithm for energy and revenue management problems was verified. Moreover, it was concluded that utilization of the BESS becomes more economical for revenue management problem as the depreciation cost of BESS decreases. 
\vspace{-0.2cm}	
\bibliographystyle{ieeetr}
\bibliography{references}

\begin{thebibliography}{10}

\bibitem{schneider2010evaluation}
K.~P. Schneider, J.~C. Fuller, F.~K. Tuffner, and R.~Singh, ``Evaluation of
  conservation voltage reduction {(CVR)} on a national level,'' tech. rep.,
  PNNL, Richland, WA (US), 2010.

\bibitem{yeo2019power}
J.~H. Yeo, P.~Dehghanian, and T.~Overbye, ``Power flow consideration of
  impedance correction for phase shifting transformers,'' in {\em 2019 IEEE
  Texas Power and Energy Conference (TPEC)}, pp.~1--6, IEEE, 2019.

\bibitem{ANSI}
``{ANSI C84.1, American National Standard For Electric Power Systems and
  Equipment—Voltage Ratings (60 Hertz)},'' 2011.

\bibitem{IEEE1547}
D.~Narang, ``{IEEE} standard for interconnection and interoperability of
  distributed energy resources with associated electric power systems
  interfaces,'' {\em IEEE Std 1547-2018 (Revision of IEEE Std 1547-2003)},
  pp.~1--138, April 2018.

\bibitem{MINLPCVR}
M.~S. Hossan, B.~Chowdhury, M.~Arora, and C.~Lim, ``Effective {CVR} planning
  with smart {DG}s using {MINLP},'' {\em North American Power Symposium
  (NAPS)}, 2017.

\bibitem{nojavan2014optimal}
S.~Nojavan, M.~Jalali, and K.~Zare, ``Optimal allocation of capacitors in
  radial/mesh distribution systems using mixed integer nonlinear programming
  approach,'' {\em Electric Power Systems Research}, vol.~107, pp.~119--124,
  2014.

\bibitem{jabr2012minimum}
R.~A. Jabr, R.~Singh, and B.~C. Pal, ``Minimum loss network reconfiguration
  using mixed-integer convex programming,'' {\em IEEE Transactions on Power
  systems}, vol.~27, no.~2, pp.~1106--1115, 2012.

\bibitem{paper1}
R.~R. Jha, A.~Dubey, C.~Liu, and K.~P. Schneider, ``Bi-level volt-var
  optimization to coordinate smart inverters with voltage control devices,''
  {\em IEEE Transactions on Power Systems}, pp.~1--1, 2019.

\bibitem{LinearEmiliano1}
A.~{Bernstein} and E.~{Dall'Anese}, ``Linear power-flow models in multiphase
  distribution networks,'' in {\em 2017 IEEE PES Innovative Smart Grid
  Technologies Conference Europe (ISGT-Europe)}, pp.~1--6, Sep. 2017.

\bibitem{gan2014convex}
L.~Gan and S.~H. Low, ``Convex relaxations and linear approximation for optimal
  power flow in multiphase radial networks,'' in {\em 2014 Power Systems
  Computation Conference}, pp.~1--9, Aug 2014.

\bibitem{LinearEmiliano2}
A.~{Bernstein}, C.~{Wang}, E.~{Dall’Anese}, J.~{Le Boudec}, and C.~{Zhao},
  ``Load flow in multiphase distribution networks: Existence, uniqueness,
  non-singularity and linear models,'' {\em IEEE Transactions on Power
  Systems}, vol.~33, pp.~5832--5843, Nov 2018.

\bibitem{teymouri2015advanced}
A.~Teymouri, S.~H. Fathi, and F.~Karbakhsh, ``An advanced hysteresis controller
  to improve voltage profile of power system with pv units: A smart grid power
  exchange framework,'' in {\em 2015 30th International Power System Conference
  (PSC)}, pp.~79--85, IEEE, 2015.

\bibitem{ostadijafari2019smart}
M.~Ostadijafari, A.~Dubey, Y.~Liu, J.~Shi, and N.~Yu, ``Smart building energy
  management using nonlinear economic model predictive control,'' {\em arXiv
  preprint arXiv:1906.00362}, 2019.

\bibitem{mohammadi2016allocation}
F.~Mohammadi, H.~Gholami, G.~B. Gharehpetian, and S.~H. Hosseinian,
  ``Allocation of centralized energy storage system and its effect on daily
  grid energy generation cost,'' {\em IEEE Transactions on Power Systems},
  vol.~32, no.~3, pp.~2406--2416, 2016.

\bibitem{rahimi2013simple}
A.~Rahimi, M.~Zarghami, M.~Vaziri, and S.~Vadhva, ``A simple and effective
  approach for peak load shaving using battery storage systems,'' in {\em 2013
  North American Power Symposium (NAPS)}, pp.~1--5, IEEE, 2013.

\bibitem{barnes2012placement}
A.~K. Barnes, J.~C. Balda, A.~Escobar-Mej{\'\i}a, and S.~O. Geurin, ``Placement
  of energy storage coordinated with smart pv inverters,'' in {\em 2012 IEEE
  PES Innovative Smart Grid Technologies (ISGT)}, pp.~1--7, IEEE, 2012.

\bibitem{yeh2012performance}
H.~Yeh, S.~Doan, and D.~Gayme, ``Performance of var controls for distribution
  lines with photovoltaic cells and batteries,'' in {\em 2012 IEEE Power and
  Energy Society General Meeting}, pp.~1--6, IEEE, 2012.

\bibitem{yeh2013battery}
H.-G. Yeh and S.~H. Doan, ``Battery placement on performance of var controls,''
  {\em arXiv preprint arXiv:1311.6199}, 2013.

\bibitem{zafar2018multi}
R.~Zafar, J.~Ravishankar, J.~E. Fletcher, and H.~R. Pota, ``Multi-timescale
  model predictive control of battery energy storage system using conic
  relaxation in smart distribution grids,'' {\em IEEE Transactions on Power
  Systems}, vol.~33, no.~6, pp.~7152--7161, 2018.

\bibitem{MFarivarLow}
M.~Farivar and S.~H. Low, ``Branch flow model: Relaxations and
  convexification—part i,'' {\em IEEE Transactions on Power Systems},
  vol.~28, pp.~2554--2564, Aug 2013.

\bibitem{kers}
W.~H. Kersting, {\em Distibution System Modeling and Analysis}.
\newblock thrid edition, CRC, 2012.

\bibitem{wei2016proactive}
T.~Wei, Q.~Zhu, and N.~Yu, ``Proactive demand participation of smart buildings
  in smart grid,'' {\em IEEE Transactions on Computers}, vol.~65, no.~5,
  pp.~1392--1406, 2016.

\end{thebibliography}

\end{document}